\documentclass[aip,graphicx]{revtex4-1}

\usepackage{amsmath, amsfonts, amssymb, mathrsfs, mathtools}
\usepackage{color, graphicx, subfigure, epstopdf}

\begin{document}

\title{Stringy Corrections to the Classical Tests of General Relativity}

\author{Christopher Frye}
\email[]{christopher.frye@knights.ucf.edu\\[-5pt]Address after August 1, 2013:\\[-5pt]Department of Physics, Harvard University,
                17 Oxford Street, Cambridge, MA 02138}
\affiliation{\mbox{Department of Physics, University of Central Florida, Orlando, FL 32816}}

\author{Costas Efthimiou}
\email[]{costas@physics.ucf.edu}
\affiliation{\mbox{Department of Physics, University of Central Florida, Orlando, FL 32816}}

\date{\today}

\begin{abstract}

String theory imposes  modifications to Einstein's equations of classical general relativity.  Consequently,  we calculate the additional
corrections to the classical tests: the perihelion precession of Mercury, the  deflection of  light rays by the sun, and the  gravitational redshift which should be present if these modified equations hold. In each case, we determine --- quite consistently with expectations --- that the stringy effects are much too small to be measured.

\end{abstract}

\maketitle

\section{Introduction}

General Relativity is a well known subject on which numerous texts (many of them well studied and popular)  exist \cite{Carroll,Landau,MTW,Pathria,Weinberg}.  None of them forgets to mention (either briefly or with many details) the tests that became the referees for the acceptance of the theory: the perihelion precession of Mercury, the  deflection of  light rays by the sun, and the  gravitational redshift --- known today as  the classical tests of  general relativity.

Although there appears to be no need to modify general relativity at the classical level, many modifications have been proposed over the years\cite{BransDicke,f(R),GSW,Gupta,CartanGravity,fromNick}, the majority of them motivated by the quest of unifying gravity with electromagnetism and the quest of a quantum theory of gravity.

Obviously, any proposed modification of general relativity will result in modified predictions for the outcomes of the classical texts. In principle, high precision in the experimental determination  of the classical tests can serve to eliminate and/or select some of such proposed modified theories.  In this paper, we investigate consequences of the modifications to gravity that string theory imposes, specifically on the classical tests of general relativity.

\section{Static Isotropic Field}

Green, Schwarz, and Witten state the schematic modification implied for Einstein's equations in their early text on string theory\cite{GSW} --- a term proportional to the square of the Riemann-curvature tensor must be added:
 \[
        R_{\mu\nu} + {\alpha' \over 2} \, R_{\mu\alpha\beta\gamma}\,R_\nu^{~\alpha\beta\gamma} ~=~
        \mathcal{O}(\alpha^{\prime \, 2})  ,
 \]
 where $\alpha'$ is the inverse string tension\cite{Zwiebach}  (we use $\hbar = c = 1$ throughout this article). Since we are concerned with the classical tests, we must use the static isotropic solution
 of these equations. However,  in another paper\cite{otherPaper}, we show that the above modification to Einstein's equations lacks  an isotropic solution. Indeed, the graviton cannot be decoupled from the dilaton\cite{BlackHoles,Gasperini} and a set of coupled equations must be used.

    More precisely, string theory imposes modifications to Einstein's field equations of general relativity, which to leading order in the string parameter $\alpha'$ take the form\cite{BlackHoles}
    \begin{align}
    \label{eq:modGrav}
        R_{\mu\nu} + 2 \, \nabla_\mu \, \nabla_\nu \, \Phi
         + \lambda \, R_{\mu\alpha\beta\gamma}\,R_\nu^{~\alpha\beta\gamma} ~=&~~ \mathcal{O}(\lambda^2)\,, \\[5pt]
        \square \Phi - (\nabla \Phi)^2 + {1 \over 4} \, R
        + {1 \over 8} \, \lambda \, R_{\alpha\beta\gamma\delta} \, R^{\alpha\beta\gamma\delta} ~=&~~
        \mathcal{O}(\lambda^2)\,.
    \label{eq:modPhi}
    \end{align}
    Here, the gravitational field is coupled to the scalar dilaton field $\Phi$\,, while $\lambda \sim 10^{-70}$ m$^2$ proportional to the string parameter $\alpha'$ as\cite{BlackHoles}
    \begin{alignat*}{2}
        \lambda &= {\alpha'/2} & \qquad &\text{in bosonic string theory}\,, \\
        \lambda &= {\alpha'/4} & \qquad &\text{in heterotic string theory}\,, \\
        \lambda &= 0           & \qquad &\text{in Type-II supersymmetric string theory}\,.
    \end{alignat*}
    The corresponding static isotropic spacetime metric outside a mass $m$ is known\cite{BlackHoles}:
    \begin{equation}
    \label{eq:fgmetric}
        ds^2 = - B(r) \, dt^2 + A(r) \, dr^2 + r^2\,d\theta^2 + r^2 \, \sin^2 \theta \, d\phi^2 ,
    \end{equation}
    with
    \begin{align}
    \label{eq:f}
        B(r) &= \left( 1-{2\,G\,m \over r} \right) \; \Bigg[ 1 - {2\lambda \over r^2} \, \left(
        {23\,r \over 24\,G\,m} + {11 \over 12} + {G\,m\over r}
        \right) \Bigg]\;, \\[15pt]
        A(r) &= \left(1-{2\,G\,m \over r} \right)^{-1} \; \Bigg[ 1 - {2\lambda \over r^2} \, \left(
        {r \over 24\,G\,m} + {7 \over 12} + {5\,G\,m \over 3\,r}
        \right) \Bigg]\,,
    \label{eq:g}
    \end{align}
    slightly altering the well-known Schwarzschild metric when $\lambda \neq 0$\,.

\section{Motion in a Static Isotropic Field}

    We are interested in using the previous result to compute the deviations from Einstein's theory that should occur in the classical tests of general relativity. These calculations deal with bodies moving in the geometry given by \eqref{eq:fgmetric}--\eqref{eq:g}. It is easiest to proceed as far as possible leaving $A(r)$ and $B(r)$ as arbitrary functions\,; we thus work with \eqref{eq:fgmetric} for now following Weinberg\cite{Weinberg}, without reference to \eqref{eq:f} or \eqref{eq:g}\,.

    To determine a particle's motion in this geometry, we need to solve the geodesic equations
    \begin{equation}
    \label{eq:geodesic}
        {d^2 x^\mu \over dp^2} + \Gamma^\mu_{~\,\nu\lambda} \, {dx^\nu \over dp} \, {dx^\lambda \over dp} = 0\,.
    \end{equation}
    By spherical symmetry, we can choose $\, \theta = \pi/2 \,$ initially, and it will remain so for all times, simplifying the equations. Since this effectively creates a $(2+1)$-dimensional problem, we are left with three components of \eqref{eq:geodesic}\,:
    \begin{align*}
        0 ~&=~ {d^2 t \over dp^2} + {B'(r) \over B(r)}\,{dt \over dp}\,{dr \over dp}\,, \\[10pt]
        0 ~&=~ {d^2 r \over dp^2} + {A'(r) \over 2\,A(r)} \, \left({dr \over dp}\right)^2 - {r \over A(r)}\,\left({d\phi \over dp}\right)^2 + {B'(r) \over 2\,A(r)}\,\left({dt \over dp}\right)^2\,, \\[10pt]
        0 ~&=~ {d^2\phi \over dp^2} + {2 \over r}\,{d\phi \over dp}\,{dr \over dp}\,.
    \end{align*}
    The $t$- and $\phi$-equations can each be written as the vanishing of a total derivative. One of the resulting constants of motion is absorbed into the parametrization $p$ of the path, while the other becomes
    \begin{equation}
    \label{eq:J}
        J = r^2\,{d\phi \over dp} = \text{const} \,.
    \end{equation}
    Manipulation of the $r$-equation then produces
    \begin{equation}
    \label{eq:EOM}
        A(r)\,\left({dr \over dp}\right)^2 + {J^2 \over r^2} - {1 \over B(r)} ~=~ - E ~=~ \text{const} ,
    \end{equation}
    where, upon computation of the interval $ds^2$, we discover that $E$ is strictly positive for massive particles and zero for massless particles. Since we are primarily concerned with the shapes of trajectories, we trade $\,dp\,$ for $\,d\phi\,$ in \eqref{eq:EOM} using \eqref{eq:J} to find
    \begin{equation}
    \label{eq:EOM2}
        {A(r) \over r^4}\,\left({dr \over d\phi}\right)^2 + {1 \over r^2} - {1 \over J^2\,B(r)} \:=\: - {E \over J^2}\,.
    \end{equation}
    Rearrangement and integration gives the equation we will use when considering the first two classical tests:
    \begin{equation}
    \label{eq:phiint}
        \phi - \phi_0 \:=\: \pm \int {dr \over r^2} \, \sqrt{A(r) \over {1 \over J^2 \, B(r)} - {E \over J^2} - {1 \over r^2}}\,.
    \end{equation}

    We are henceforth concerned with situations in which the gravitational potential $Gm/r$ is quite small. Indeed, just outside our Sun we have $\,Gm/r \simeq 2 \times 10^{-6}$\,. Thus, keeping only the leading-order corrections to the Schwarzschild metric, we will use
    \begin{equation}
    \label{eq:ourmetric}
        ds^2 \:=\: - \left( 1 - {2\,G\,m \over r} - {23 \, \lambda \over 12 \, G\,m\,r} \right) \, dt^2 \:+\:
        \left(  1 + {2\,G\,m \over r} - {\lambda \over 12\,G\,m\,r} \right) \, dr^2 \:+\: \cdots
    \end{equation}
    instead of \eqref{eq:f} and \eqref{eq:g}. This redefines the functions $A(r)$ and $B(r)$ to substitute in \eqref{eq:phiint}.

\section{Consequences for the Classical Tests}

    We begin by calculating the additional deflection of light by a star that should occur due to string theory. This is a scattering problem in which we are interested in the total change in $\phi$ as a photon approaches the Sun from infinity and returns to infinity. To calculate this we use  equation \eqref{eq:phiint} with $E=0$\,. We can find $J$ from \eqref{eq:EOM2} since $dr/d\phi = 0$ when $r = r_0$, the minimum distance between the photon and the star's center. Since the geodesic equation \eqref{eq:geodesic} is invariant under ``time reversal'' $p \mapsto -p$, the trajectory of the photon is also symmetric around its point of closest approach to the Sun. Hence, we only need to calculate the change in $\phi$ over half the path, then double it. If the Sun did not deflect light at all, we would expect a change in $\phi$ of precisely $\pi$ radians. Putting all these considerations together, the total deflection of a photon as it passes the Sun is
    \[
        \Delta \phi = 2 \, \left| ~\int_{r_0}^\infty {dr \over r} \sqrt{A(r) \over \left( {r \over r_0} \right)^2 \, \left( {B(r_0) \over B(r)} \right) - 1} ~ \right| - \pi\,,
    \]
    into which we must insert $A(r)$ and $B(r)$ from \eqref{eq:ourmetric}.
    Consider expanding this integrand in powers of $\lambda$. The $\lambda^0$ term would integrate to give the classical result of
    \[
        \Delta \phi = {4\,m\,G \over r_0}\,;
    \]
    this is correct to zeroth order in $\lambda$ and first order in the potential. For light passing near the surface of our Sun, this gives $1.75''$. Now integrating the $\lambda^1$ term gives the small deviation $\delta\phi$ from the classical result. To leading order in the potential,
    \begin{equation}
    \label{eq:answerdefl}
        \delta\phi ~=~ {\lambda \over 12\,G\,m\,r_0} \int_{r_0}^\infty {\big(23\,r^2 - r\,r_0 - r_0^2 \big)\,dr \over (r+r_0) \, \sqrt{\left({r \over r_0}\right)^2 - 1}} ~=~ {11 \, \lambda \over 6\,G\,m\,r_0}\,.
    \end{equation}
    Since $\delta\phi$ is positive, the total deflection is slightly greater than the classical deflection. For light just grazing our Sun, $\delta\phi$ is of order $10^{-82}$ radians. Therefore, stringy effects should shift the reception point of the photon on Earth by an additional $10^{-71}$ meters. We will comment on the size of these corrections in Section \ref{sec:conc}, after considering all phenomena.

    We proceed to calculate in a similar way the additional precession of planetary orbits due to string theory. We determine $E$ and $J$ from \eqref{eq:EOM2} using the fact that $dr/d\phi = 0$ when $r = r_\pm$, the radii corresponding to the aphelion and perihelion of the planet's orbit. As above, we integrate over only half the path length and double the result. In this case we subtract $2\pi$ from the integral, because deviation from $2\pi$ implies a precession in the orbit-ellipse. Using these ideas, we determine from \eqref{eq:phiint} that the total precession $\Delta \phi$ of an orbit-perihelion is given by
    \[
        2 \, \left| ~\int_{r_-}^{r_+} {\sqrt{A(r)} \; dr \over r^2} \left/ \sqrt{{r_-^2 \, \big[ 1/B(r) \, - \, 1/B(r_-) \big] - r_+^2 \, \big[ 1/B(r) - 1/B(r_+) \big] \over r_+^2\,r_-^2 \, \big[ 1/B(r_+) - 1/B(r_-) \big]} - {1 \over r^2}} \right. ~\right| - 2\pi
    \]
    with substitution for $A(r)$ and $B(r)$ from \eqref{eq:ourmetric}.
    Again, consider expanding this integrand in powers of $\lambda$. The $\lambda^0$ term would integrate to give the classical result of
    \[
        \Delta\phi = 3\,\pi\,m\,G\,\left({1 \over r_+} + {1\over r_-}\right) ~ {\text{radians} \over \text{revolution}}\,;
    \]
    this is correct to zeroth order in $\lambda$ and first order in the potential. For Mercury orbiting our Sun, this gives $43.03''$ per century. Now integrating the $\lambda^1$ term gives the small deviation $\delta\phi$ from the classical result. To leading order in the potential,
    \begin{equation}
    \label{eq:answerprec}
        \delta\phi ~=~ {\lambda \over 12\,G\,m\,\sqrt{r_- \, r_+}} \int_{r_-}^{r_+} {(23\,r\,r_- + 23\,r\,r_+ - r_-\,r_+)\,dr \over r^2 \, \sqrt{(r_+ - r)\,(r - r_-)}}
        ~=~ {15\,(r_- + r_+) \, \pi \, \lambda \over 8 \, G\,m\,r_+\,r_-}\,.
    \end{equation}
    Since $\delta\phi$ is positive, string theory implies a slightly faster precession than classical general relativity. For Mercury, $\delta\phi$ is of order $10^{-83}$ radians per revolution. It would take roughly $10^{26}$ times the age of the universe for Mercury to advance an extra Planck-length due to stringy effects.

    Finally, we calculate the gravitational redshift of light as it travels away from a massive body. If we sit at $x_1$ and receive radiation from identical atomic transitions (say) occurring at distinct points $x_1$ and $x_2$\,, then the frequencies we measure will obey
    \[
        {\nu_2 \over \nu_1} = \sqrt{g_{00}(x_2) \over g_{00}(x_1)}\,.
    \]
    To zeroth order in $\lambda$, we find the classical result
    \[
        {\Delta \nu \over \nu} = \phi(x_2) - \phi(x_1)
    \]
    where $\phi(x)$ here is the gravitational potential at $x$\,; this result is correct to first order in the potential. For light received on Earth from atomic transitions on the Sun, this becomes roughly $-2 \times 10^{-6}$\,; here, the negative result signifies the decrease in frequency as light travels away from the Sun. To lowest order in the potential, the $\lambda^1$ term in this expression gives the small correction
    \begin{equation}
    \label{eq:answerred}
        {\delta \nu \over \nu} = - {23\lambda\over 24 \,G m} \, \left( {1 \over r} - {1 \over R} \right)
    \end{equation}
    due to string theory; here $r$ is the radius of the Sun, while $R$ is the distance between Earth's surface and the Sun's center. The fact that $\,\delta \nu < 0 \,$ shows that, once again, the extra shift due to string theory is in the same direction as the classical effect. From Sun to Earth, this result becomes of order $10^{-82}$\,.

\section{Analysis \& Conclusion}
\label{sec:conc}

    We have calculated the small changes in the results of three classical tests of general relativity that we should expect if the precise stringy equations are indeed true. The results found in each case are unimaginably small. Even if, instead of near our Sun, we let these phenomena occur in the vicinity of a distant black hole, this will not change our chances at measuring any of these effects.

    Indeed, let us consider maximizing $\delta \phi$ in the deflection of light. If we look at a ray that grazes the Schwarzschild sphere
    of a black hole where $Gm/r_0 = 1/2\,$, then
    \begin{equation}
    \delta\phi ~=~  {11 \, \lambda \over 3\,r^2_0}\,.
    \end{equation}
    If we consider a black hole the size of a nucleus, $\,r_0 \sim 10^{-15}\,$,  and let the scattering occur near the edge of the observable universe $10^{26}$ m away , this would still require us to distinguish $10^{-14}$ m between reception points on Earth. In observing such a distant event, pushing uncertainties so low appears impossible.  In addition, when  considering micro black holes, leaving aside quantum mechanics is not technically correct.  However, to entertain our reader we ignore this difficulty here. If stable mini black holes are produced with $\,r_0 \sim 10^{-35}\,$, then we get some hope to see an experimental verification.

    Since both the precession-of-perihelion and gravitational-redshift phenomena also have $\,1/Gmr\,$-\,dependencies but do not benefit from observation a great distance away, the situation is much worse for the other two tests. We must thus search elsewhere for experimental tests of string theory until, perhaps,  tiny mini black holes are observed or produced.

    String cosmologists are most optimistic when it comes to experimentally testing the theory\cite{Baumann,Kallosh,Tye}. Many hope to draw conclusions about string theory's validity by both investigating the dynamics of the early universe through the cosmic microwave background radiation and also determining how high-energy physics can produce ``peculiar'' low-energy effects, e.g.~the large size of our universe\cite{McAllister}.

\begin{acknowledgments}

    CF expresses his deepest gratitude to his undergraduate advisor CE for suggesting these calculations and providing guidance during this project. Both authors thank  Tristan H\"ubsch for reading the initial manuscript  and providing feedback.

\end{acknowledgments}


\end{document}